\documentclass[aps,prc]{revtex4}
\usepackage{epsfig}
\newcommand{\vsig}{\mbox{\boldmath$\sigma$\unboldmath}}
\newcommand{\veps}{\mbox{\boldmath$\epsilon$\unboldmath}}

\newcommand{\be}{\begin{equation}}
\newcommand{\ee}{\end{equation}}
\newcommand{\bea}{\begin{eqnarray}}
\newcommand{\eea}{\end{eqnarray}}
\newcommand{\bean}{\begin{eqnarray*}}
\newcommand{\eean}{\end{eqnarray*}}

\newcommand{\gapproxeq}{\lower
.7ex\hbox{$\;\stackrel{\textstyle >}{\sim}\;$}}
\newcommand{\lapproxeq}{\lower
.7ex\hbox{$\;\stackrel{\textstyle <}{\sim}\;$}}

\begin{document}

\title{\bf Vector meson photoproduction studied in its radiative decay channel}

\author{Qiang Zhao$^1$\footnote{Email address: qiang.zhao@surrey.ac.uk},
J.S. Al-Khalili$^1$\footnote{Email address: j.al-khalili@surrey.ac.uk},
and P.L. Cole$^2$\footnote{Email address: cole@physics.isu.edu} }
\affiliation{1) Department of Physics, University of Surrey,
Guildford, Surrey GU2 7XH, United Kingdom}
\affiliation{2) Department of Physics, Idaho State University,
Pocatello, Idaho 83209, USA}

\date{\today}

\begin{abstract}
We provide an analysis
of vector meson photoproduction 
in the channel of the vector meson decaying into a pseudoscalar meson 
plus a photon, i.e.  $V\to P\gamma$. 
It is shown that non-trivial kinematic correlations arise from the
measurement of the $P\gamma$ angular distributions 
in the overall c.m. system in comparison with those 
in the vector-meson-rest frame. 
In terms of the vector meson density matrix elements, 
the implication of such kinematic correlations in the measurement
of polarization observables is discussed. 
For the
$\omega$ meson production,
due to its relatively large branching ratios for $\omega\to\pi^0\gamma$,
additional events from this channel may
enrich the information about the reaction mechanism
and improve the statistics of the recent
measurement of polarized beam asymmetries by the GRAAL Collaboration. 
For $\phi\to \eta\gamma$, $\rho\to \pi\gamma$, and $K^*\to K\gamma$,
 we expect that additional information about the spin structure of the vector meson 
production vertex can be derived.

\end{abstract}

\maketitle


\section{Introduction}

Direct experimental evidence for baryon resonance couplings into
vector meson has been searched for
in vector meson photoproduction reaction 
in recent years~\cite{saphir,graal,e91-016,clas}.
One of the motivations behind this effort
is to find ``missing resonances", which are
predicted by the nonrelativistic constituent quark model (NRCQM)
but not found in $\pi N$ scatterings~\cite{isgur,capstick-roberts}.
The study of vector meson photoproduction at large angles 
near threshold is particularly useful
due to the relatively small contributions from the background processes,
such as {\it t}-channel natural and unnatural parity exchanges.

Various experimental projects have been and are being carried out
at ELSA, JLab, ESRF, and Spring-8. In addition to the cross sections,
polarization observables, which are more sensitive
to resonance excitations, will also be measured using newly developed
experimental techniques~\cite{didelez,cole}. This is of great importance
for the purpose of disentangling the {\it s}-channel resonance
excitations in the reaction.
Since a large number of degrees of freedom will be involved and
due to the lack of dynamic information in such a non-perturbative region,
one, in principle, needs a complete set of
measurements of all independent spin polarization observables to obtain
sufficient information about resonance excitations
and their couplings to the meson and nucleon~\cite{tabakin}.

The polarization observables and
density matrix elements are essentially equivalent languages
that connect the theoretical phenomenologies
with the experimental observables. 
The density matrix elements, as the interference
between the independent transition amplitudes, can be 
measured in the
final state vector meson decay distributions. 
They can be directly compared with theoretical calculations, 
and hence give access to dynamical information about the transition mechanism. 
For the vector meson photoproduction reaction,
A number of analyses have been carried out in the literature.
Density matrix elements for polarized photon beams
were discussed by Schilling, Seyboth and Wolf~\cite{schilling}. 
More recently, Pichowsky, \c{S}avkl\i~and Tabakin~\cite{tabakin}
studied various relations between polarization observables
and helicity amplitudes.  In Ref.~\cite{Kloet,Kloet-2},
Kloet {\it et al.} investigated inequality constraints on
the density matrix elements defined for the produced vector meson
with unpolarized or linearly polarized photon beams.
They also explored relations between vector meson decay distributions
in the vector-meson rest frame and the overall $\gamma$-$N$ frame, 
which can be extended to the analysis of other decay 
channels.

Another aspect relevant to the derivation of the density
matrix elements is the dynamics for the produced vector meson
coupling to the detected particles.
For instance, $\omega$ meson is usually detected in $\omega\to 
\pi^+\pi^-\pi^0$
and $\rho^0$ in $\rho^0\to\pi^+\pi^-$. The angular distributions
for vector meson decays into spinless particles were studied in
Ref.~\cite{schilling}. However, their decays into photon and 
pseudoscalar meson, $V\to P\gamma$, 
have not been discussed. In this work, we will develop the
formalism for vector meson decay into
photon and pseudoscalar meson, e.g. $\omega\to \pi^0\gamma $, 
$\phi\to\eta\gamma$, $\rho\to  \pi\gamma$, and $K^*\to K\gamma$. 
We will show that
additional information about the vector meson
production mechanism can be obtained from measurements of those
final states.
This analysis has an
advantage in $\omega$ meson photoproduction, of which the branching
ratio of $\omega\to  \pi^0\gamma$ is about 8.5\%~\cite{pdg2004}. 
As an isoscalar meson, only nucleon resonances can contribute 
in its {\it s}- and {\it u}-channel production.
This reduces significantly the number of excited states in
the corresponding reaction channels. In this sense, the additional
information from $\omega\to \pi^0\gamma $ should be useful for
constraining the nucleon resonances in $\gamma N\to \omega N$.
Nevertheless, since the $\omega$ decays are dominantly via $\omega\to \pi^+\pi^-\pi^0$, 
and $\pi^+\pi^-$ could form a configuration of $J^P=1^{-}$ as a photon, 
it is interesting to compare these two distributions and gain some insights 
into the sensitivities of the vector meson production mechanism to 
these two decay channels in a polarization measurement. This possibility 
is enhanced in the comparison between $V\to PP$ and $V\to P\gamma$.  
For $\phi\to\eta\gamma$, the branching ratio is still sizeable. 
In comparison with the dominant decay channel of $\phi\to K\bar{K}$, 
the decay distribution of $\phi\to\eta\gamma$ contains different spin structure 
information; hence should be more sensitive to the production mechanism. 
Similar feature applies to the $\rho\to\pi\gamma$ and 
$K^*\to K\gamma$ in $\rho$ and $K^*$ meson photoproduction.

In this paper, the analysis of 
the decay channel $V\to P\gamma$, will be presented in Sect. II.
In Sect. III, we discuss the kinematic correlations 
arising from this decay channel and analyze their influence 
on the measurements of spin observables 
in the the overall $\gamma N$ c.m. system. Comparison with the results
derived in the vector-meson-rest frame will then be made. 
A summary will be given in Sec. IV. 
To make it convenient for readers, 
analyses of the spin observables in terms of the 
density matrix elements of the polarized particles are included
in the Appendix.

\section{Density matrix elements for $V\to P\gamma$}

In the overall c.m. system, the invariant amplitude is defined as
\begin{equation}
T_{\lambda_v\lambda_f,\lambda_\gamma\lambda_i}\equiv
\langle  {\bf q},\lambda_v; {\bf P}_f, \lambda_f | T
|{\bf k}, \lambda_\gamma; {\bf P}_i, \lambda_i\rangle
\to
\langle \lambda_v \lambda_f | T |\lambda_\gamma \lambda_i\rangle ,
\end{equation}
where ${\bf k}$ and ${\bf q}$ are momenta of the initial photon and final
state vector meson, respectively; Momenta ${\bf P}_i=-{\bf k}$
and ${\bf P}_f=-{\bf q}$ are for the initial and final state nucleons,
respectively; $\lambda_\gamma$ ($=\pm 1$), $\lambda_v$ ($=0,\pm 1$),
$\lambda_i$ ($=\pm 1/2$) and $\lambda_f$ ($=\pm 1/2$) are helicities of the
photon, vector meson and initial and final state nucleons, respectively.

The amplitudes for the two photon polarizations $\lambda_\gamma=\pm 1$
are not independent
of each other. They are connected by the Jacob-Wick parity relation:
\begin{equation}
\label{jacob-wick}
\langle  \lambda_v \lambda_f|T| \lambda_\gamma\lambda_i\rangle
= (-1)^{(\lambda_v-\lambda_f)-(\lambda_\gamma-\lambda_i)}
\langle -\lambda_v -\lambda_f|T|-\lambda_\gamma -\lambda_i\rangle.
\end{equation}

We first consider the vector meson decay in the overall $\gamma$-$N$ c.m. 
system.
The decay of vector meson ($J^{PC}=1^{-}$) into a pseudoscalar meson 
($J^{P}=0^{-}$) and photon ($J^{PC}=1^{--}$), i.e.  
$V\to P\gamma$, is described by the effective Lagrangian
\begin{equation}
\label{vpr}
{\cal L}_{VP\gamma}=\frac{e g_V}{M_V}
\epsilon_{\alpha\beta\gamma\delta}\partial^\alpha A^\beta
\partial^\gamma V^\delta P \ ,
\end{equation}
where $g_V$ is the coupling constant and
$\epsilon_{\alpha\beta\gamma\delta}$ is the Levi-Civita tensor;
$V^\delta$ and $P$ here denote the vector and pseudoscalar meson fields,
and $A^\beta$ is the photon field; This effective Lagrangian 
is the only Lorentz-covariant one for the $VP\gamma $ coupling 
at leading order. 

We assume that the fields couple as elementary ones. 
Therefore, the momentum-dependence of $g_V$ can be neglected, i.e. 
$g_V=g_V(0)$. 
This assumption depends on the average radius ${\langle r_q^2 \rangle}^{1/2}$ 
of the vector meson, for which we phenomenologically 
refer to the average disctance between 
the quark and antiquark within the meson. 
As a simple estimate for the ground state mesons of which 
the spatial wavefunctions are spherical, 
the particle size effects appear as the quadratic term in the form factor 
$1-k_q^2 \langle r_q^2 \rangle/6$, where $k_q\simeq 380$ MeV are the typical 
constituent quark momenta. For a typical size of 
${\langle r_q^2 \rangle}^{1/2}\sim 0.5$ fm for the mesons, 
a correction of about 15\% seems to be needed. 
However, this will not affect the general feature of the subsequent analyses 
due to two reasons:
i) As shown by Becchi and Morpurgo~\cite{bm}, in the limit 
of $k_q=0$, i.e. $g_V(k_q^2)=g_V(0)$, the calculations for both limits of 
$M_V=M_P(exp)$ and $M_P=M_V(exp)$ 
give the same results which are consistent with the experimental data. 
This suggests that the momentum dependence of $g_V$ is not significant. 
ii) Even though $g_V(k_q^2)\neq g_V(0)$, 
it will only lead to some percentages of descrepancies in the event countings
between the calculation of Eq.~(\ref{vpr}) and experimental statistics, 
while the decay distributions will not change. 
In this sense, we can neglect the the momentum dependence of the $g_V$ 
at this moment, but focus on the study of kinematic correlations 
arising from the measurement of the $P\gamma$ angular distributions. 

It is worth noting that the vector meson size effects may become 
non-negligible at the production vertices for $\gamma N\to V N$, and 
careful considerations of the vector meson coupling form factors 
are crucial. However, this is independent of our motivation here. 
As mentioned in the Introduction, we are interested in possible 
experimental measurements of the angular distributions of final state particles, 
through which dynamical information about the production mechanism 
(contained in the density matrix elements) 
can be extracted.

\begin{figure}
\begin{center}
\epsfig{file=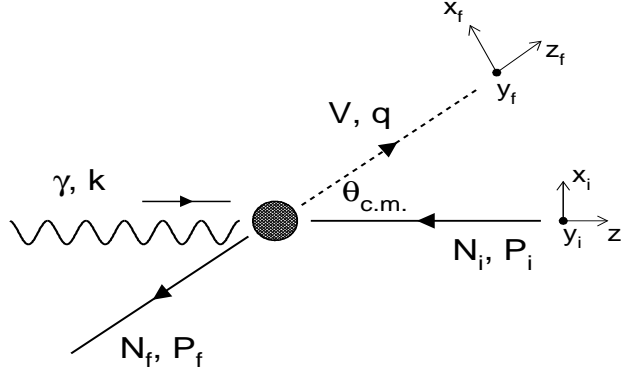, width=11cm,height=9.cm}
\caption{Kinematics for $\gamma N_i\to V N_f$ in the overall c.m. system.
}
\protect\label{fig:(1)}
\end{center}
\end{figure}

\begin{figure}
\begin{center}
\epsfig{file=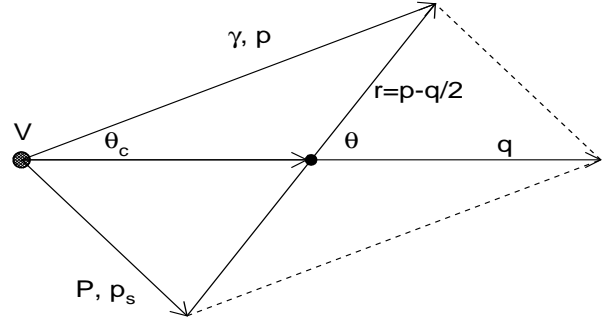, width=11cm,height=9.cm}
\caption{Kinematics for $V\to P \gamma$ in the overall c.m. system where the 
vector meson has momentum ${\bf q}$.  
The angle $\theta_c$
denotes the decay direction of the photon, while 
$\theta$ is the angle between ${\bf q}$ and the momentum difference 
between the photon and pseudoscalar meson. 
}
\protect\label{fig:(2)}
\end{center}
\end{figure}

As illustrated in Fig.~\ref{fig:(1)}, we define the $z_f$ axis along the
three moment of the vector meson, the $y_f$ axis is normal to the production
plane defined by ${\bf k}\times {\bf q}$, and hence the $x_f$ direction is
determined by $y_f\times z_f$.
Since the vector meson three momentum $|{\bf q}|$ is determined by the
c.m. energy $W$, the three momentum of the final state photon in the
overall c.m. system can be expressed as ${\bf p}={\bf q}/2 +{\bf r}$,
where ${\bf r}$ is the three momentum of the photon in the vector meson
rest frame.
The $z_f$ axis also defines the $z$ direction of the decay co-ordinates, 
while the polar and azimuthal angles $\theta$ and $\phi$ are defined 
by the momentum difference between the final-state photon 
and pseudoscalar meson, ${\bf p}-{\bf p_s}=2{\bf r}$. 
On the other hand, for the description of vector meson decay, 
it is convenient to select the decay angles, $\theta_c$ and $\phi_c$, 
which are the polar and azimuthal angles of the flight direction of the photon
in the overall c.m. system with respect to $z_f$. As shown in Fig.~\ref{fig:(2)}, 
we have, 
\begin{eqnarray}
p_0\sin\theta_c &= & |{\bf r}|\sin\theta, \nonumber\\  
p_0\cos\theta_c &= & |{\bf r}|\cos\theta + |{\bf q}|/2, \nonumber\\
\phi_c &= &\phi \ , 
\end{eqnarray}
where $p_0=|{\bf p}|$ is the energy of the final-state photon. 

The decay amplitude of the vector meson with transverse polarization
can thus be expressed as:
\begin{eqnarray}
\label{m-trans}
\langle \Lambda_\gamma ; \theta_c,\phi_c |M|\lambda_v=\pm 1 \rangle &=&
C\sqrt{\frac{3}{8\pi}} 
\frac{p_0(q^0\Lambda_\gamma +|{\bf q}|\lambda_v)}{M_V} 
\veps_\gamma\cdot\veps_v \nonumber\\
&=& C\sqrt{\frac{3}{8\pi}} 
\frac{p_0(q^0\Lambda_\gamma +|{\bf q}|\lambda_v)}{M_V}
(-1)^{\lambda_v}D^{1*}_{-\lambda_v\Lambda_\gamma}(\phi_c,\theta_c,-\phi_c) , 
\end{eqnarray}
and the amplitude with longitudinal polarization is:
\begin{eqnarray}
\langle \Lambda_\gamma ; \theta_c,\phi_c |M|\lambda_v=0 \rangle &=&
C\sqrt{\frac{3}{8\pi}}
|{\bf p}| \Lambda_\gamma \veps_\gamma\cdot\hat{\bf q} \nonumber\\
&=&  C\sqrt{\frac{3}{8\pi}}|{\bf p}| \Lambda_\gamma
D^{1*}_{0\Lambda_\gamma}(\phi_c,\theta_c,-\phi_c) ,
\end{eqnarray}
where $\Lambda_\gamma=\pm 1$ and $\lambda_v=0, \ \pm 1$ are the helicities
of the photon and vector meson in $V\to P \gamma$, respectively.
The coefficient $C=eg_V$ is the coupling constant.

The Wigner rotation functions follow
the convention of Rose~\cite{rose}:
\begin{equation}
\label{rotation}
D_{MN}^I(\alpha,\beta,\gamma)=e^{-i(M\alpha+N\gamma)}d_{MN}^I(\beta), 
\end{equation}
where $\alpha$, $\beta$, and $\gamma$ are Euler angles for the rotations 
of a vector.

Similar to the case of vector meson decaying into spinless 
particles~\cite{schilling},
we can express the angular distribution of $V \to P\gamma$
in terms of the vector meson density matrix elements
$\rho_{\lambda_v\lambda_v^\prime}$:
\begin{eqnarray}
\label{distr-pi-gamma}
\frac{d N}{d\cos\theta d\phi} 
&= & W(\cos\theta_c,\phi_c, \rho) \nonumber\\
&= & W_{TT}(\cos\theta_c,\phi_c, \rho)+ 
W_{LL}(\cos\theta_c, \phi_c, \rho)
+W_{TL}(\cos\theta_c, \phi_c, \rho) ,
\end{eqnarray}
where $\theta_c$ and $\phi_c$ are functions of $\theta$ and $\phi$; 
and the latter ones are  
the variables of the differential distributions on the left-hand side; 
the subscription $TT$, $LL$ and $TL$ denote the interfering 
distribution functions
between the transverse-transverse, longitudinal-longitudinal and 
transverse-longitudinal
vector meson polarizations. 

They have the following expressions in the 
overall c.m. system:
\begin{eqnarray}
W_{TT}(\cos\theta_c,\phi_c, \rho)
&=&
\frac{1}{\sigma_0}
\sum_{\lambda_v, \lambda_v^\prime=\pm 1; \Lambda_\gamma, 
\Lambda_\gamma^\prime}
\langle\Lambda_\gamma; \theta,\phi | M |\lambda_v\rangle
\rho_{\lambda_v\lambda_v^\prime}
\delta_{\Lambda_\gamma \Lambda_\gamma^\prime}
\langle \lambda_v^\prime| M^\dag |\Lambda_\gamma^\prime; \theta,\phi\rangle 
\nonumber\\
&=& \frac{C^2}{\sigma_0} \frac{3}{8\pi}\sum_{\lambda_v, \lambda_v^\prime=\pm 1; \Lambda_\gamma}
\frac{p_0(q_0\Lambda_\gamma +|{\bf q}|\lambda_v)}{M_V}
(-1)^{\lambda_v}
D^{1*}_{-\lambda_v\Lambda_\gamma}(\phi_c,\theta_c,-\phi_c)
\rho_{\lambda_v\lambda_v^\prime}(V)
\nonumber\\
&&\times (-1)^{\lambda_v^\prime}
D^1_{-\lambda_v^\prime\Lambda_\gamma^\prime}(\phi_c,\theta_c,-\phi_c)
\frac{p_0(q_0\Lambda_\gamma^\prime +|{\bf q}|\lambda_v^\prime)}{M_V}
\nonumber\\
&=& \frac{C^2}{\sigma_0} \frac{3}{8\pi}
\frac{p_0^2}{2}
\left\{ \big[ {\cal F}_1^2(\theta_c)
+{\cal F}_2^2(\theta_c)\big]
(\rho_{11} +\rho_{-1-1}) + \sin^2\theta_c \big[ e^{-2i\phi_c}\rho_{1-1} + 
e^{2i\phi_c}\rho_{-11}\big]\right\},
\end{eqnarray}
and
\begin{eqnarray}
W_{LL}(\cos\theta_c, \phi_c, \rho)
&=& \frac{C^2}{\sigma_0} \frac{3}{8\pi}
\sum_{\lambda_v= \lambda_v^\prime=0; \Lambda_\gamma, \Lambda_\gamma^\prime}
|{\bf p}|^2 \Lambda_\gamma\Lambda_\gamma^\prime
D^{1*}_{0\Lambda_\gamma}(\phi_c,\theta_c,-\phi_c)
\delta_{\Lambda_\gamma \Lambda_\gamma^\prime}\rho_{00}(V)
D^1_{0\Lambda_\gamma^\prime}(\phi_c,\theta_c,-\phi_c)\nonumber\\
&=&\frac{C^2}{\sigma_0} \frac{3}{8\pi} |{\bf p}|^2 \sin^2\theta_c \rho_{00} ,
\end{eqnarray}
and
\begin{eqnarray}
W_{TL}(\cos\theta_c, \phi_c, \rho)
&=&  \frac{C^2}{\sigma_0} \frac{3}{8\pi}
\sum_{\lambda_v, \lambda_v^\prime\neq 0; \Lambda_\gamma}
\frac{p_0 |{\bf p}| \Lambda_\gamma }{M_V}\left\{
(q_0\Lambda_\gamma + |{\bf q}|\lambda_v)
(-1)^{\lambda_v} D^{1*}_{-\lambda_v\Lambda_\gamma}(\phi_c,\theta_c,-\phi_c)
\rho_{\lambda_v 0}
D^1_{0\Lambda_\gamma}(\phi_c,\theta_c,-\phi_c)\right.\nonumber\\
&& \left. +(q_0\Lambda_\gamma + |{\bf q}|\lambda_v^\prime)
(-1)^{\lambda_v^\prime}
D^{1*}_{0\Lambda_\gamma}(\phi_c,\theta_c,-\phi_c)
\rho_{0\lambda_v^\prime}
D^1_{-\lambda_v^\prime\Lambda_\gamma}(\phi_c,\theta_c,-\phi_c)\right\} \nonumber\\
&=& -  \frac{C^2}{\sigma_0} \frac{3}{8\pi}
\frac{ p_0^2 \sin\theta_c}{2\sqrt{2}}
({\cal F}_1(\theta_c)-{\cal F}_2(\theta_c) )
\big[e^{-i\phi_c} (\rho_{10}-\rho_{0-1}) 
+e^{i\phi_c}(\rho_{01}-\rho_{-10})\big] ,
\end{eqnarray}
where 
\begin{eqnarray}
{\cal F}_1(\theta_c) & \equiv & \frac{1}{M_V} (q_0+|{\bf q}|)(1-\cos\theta_c) ,\nonumber\\
{\cal F}_2(\theta_c) & \equiv & \frac{1}{M_V} (q_0-|{\bf q}|)(1+\cos\theta_c)
\end{eqnarray}
are functions of $\theta_c$, and hence introduce kinematic correlations 
to the coordinate transformation. 
The $\delta$ function $\delta_{\Lambda_\gamma\Lambda_\gamma^\prime}$
implies the sum over the final-state unpolarized photons.
The factor $\sigma_0=C^2|{\bf r}|^2$ is the normalization factor 
for $V\to P\gamma$, 
and is proportional to the decay width for $V\to P\gamma$ 
in the vector meson rest frame. 
Note that $W(\cos\theta_c,\phi_c, \rho)$ is coordinate-dependent 
due to the Lorentz transformation 
from the vector meson rest frame to the overall 
c.m. system.

The density matrix elements of the vector meson are independent of 
the co-ordinate frame selection for the vector meson decays. They are related
to the initial photon polarizations in the overall c.m. system via:
\begin{equation}
\label{rho-v}
\rho_{\lambda_v\lambda_v^\prime}(V)=
\frac{1}{N} \sum_{\lambda_f\lambda_\gamma\lambda_i\lambda_\gamma^\prime}
T_{\lambda_v\lambda_f,\lambda_\gamma\lambda_i}
\rho_{\lambda_\gamma\lambda_\gamma^\prime}(\gamma)
T_{\lambda_v^\prime\lambda_f,\lambda_\gamma^\prime\lambda_i}^* \ ,
\end{equation}
where $N\equiv \frac 12 \sum_{\lambda_v\lambda_f\lambda_\gamma\lambda_i}
|T_{\lambda_v\lambda_f,\lambda_\gamma\lambda_i}|^2 $
is the normalization factor and is double
the unpolarized cross section ignoring the phase space factor.

The polarized-photon density matrix element $\rho(\gamma)$
is defined as~\cite{schilling}:
\begin{equation}
\label{photon-pol}
\rho(\gamma)=\frac 12 ( I_\gamma +\vsig\cdot{\bf P}_\gamma) ,
\end{equation}
where $\vsig$ is the Pauli
matrix for the photon's two independent polarizations;
${\bf P}_\gamma$ determines both the degree of polarization (via its
magnitude $P_\gamma$) and the polarization direction.
For linearly polarized photons,
with $\Phi$ denoting the angle between the polarization vector of the
photon $(\cos\Phi, \ \sin\Phi, \ 0)$
and the production plane ($x_i$-$z_i$ plane),
one has
${\bf P}_\gamma=P_\gamma (-\cos 2\Phi, \ -\sin 2\Phi, \ 0)$.
For the circularly polarized photons,
the polarization vector is along the $z_i$ axis and hence 
${\bf P}_\gamma=P_\gamma (0, \ 0, \ \lambda_\gamma)$ with 
$\lambda_\gamma=\pm 1$.

Substituting Eq.~(\ref{photon-pol}) into (\ref{rho-v}),
we obtain the familiar form of the vector meson density matrix elements~\cite{schilling}:
\begin{eqnarray}
\label{dens-pol-ph}
\rho^0_{\lambda_v\lambda_v^\prime}&=
&\frac{1}{2N}\sum_{\lambda_\gamma\lambda_f\lambda_i}
T_{\lambda_v
\lambda_f,
\lambda_\gamma\lambda_i}
T^*_{\lambda_v^\prime\lambda_f, \lambda_\gamma\lambda_i},
\nonumber\\
\rho^1_{\lambda_v\lambda_v^\prime}&=
&\frac{1}{2N}\sum_{\lambda_\gamma\lambda_f\lambda_i}
T_{\lambda_v\lambda_f,
-\lambda_\gamma\lambda_i}
T^*_{\lambda_v^\prime\lambda_f, \lambda_\gamma\lambda_i},
\nonumber\\
\rho^2_{\lambda_v\lambda_v^\prime}&=
&\frac{i}{2N}\sum_{\lambda_\gamma\lambda_f\lambda_i}
\lambda_\gamma
T_{\lambda_v\lambda_f, -\lambda_\gamma\lambda_i}
T^*_{\lambda_v^\prime\lambda_f,
\lambda_\gamma\lambda_i},\nonumber\\
\rho^3_{\lambda_v\lambda_v^\prime}&=
&\frac{i}{2N}\sum_{\lambda_\gamma\lambda_f\lambda_i}
\lambda_\gamma
T_{\lambda_v\lambda_f, \lambda_\gamma\lambda_i}
T^*_{\lambda_v^\prime\lambda_f,
\lambda_\gamma\lambda_i} \ .
\end{eqnarray}

The decomposition of the vector meson decay distribution in terms of the
initial photon polarizations thus allows us to express Eq.~(\ref{distr-pi-gamma})
as
\begin{equation}
\label{distr-decomp}
W(\cos\theta_c,\phi_c,\rho)=W^0(\cos\theta_c,\phi_c, \rho^0)+\sum_{\alpha=1}^3
P_\gamma^\alpha W^\alpha(\cos\theta_c,\phi_c, \rho^\alpha) \ ,
\end{equation}
where $W^0$ denotes the distribution with the unpolarized photons,
and $W^{1,2}$ denote those with the linearly polarized photons, and 
$W^3$ with circularly polarized photons.

With the Jacob-Wick parity relation [Eq.~(\ref{jacob-wick})] and the requirement that the density matrix 
elements must be Hermitian:
\begin{equation}
\rho^\alpha_{\lambda_v\lambda_v^\prime}
=\rho^{\alpha *}_{\lambda_v^\prime\lambda_v} , 
\end{equation}
the decomposed distributions can be obtained: 
\begin{eqnarray}
\label{w-0-cm}
W^0(\cos\theta_c,\phi_c, \rho^0)& = &
\frac{3}{8\pi} \frac{C^2p_0^2}{\sigma_0} \left\{ \sin^2\theta_c
\rho^0_{00} 
+ \frac{1}{2}\big[ {\cal F}_1^2(\theta_c)
+{\cal F}_2^2(\theta_c) \big] \rho^0_{11} 
\right.\nonumber\\
&&\left. + \sin^2\theta_c\cos 2\phi_c\rho^0_{1-1}
+\sqrt{2} [ {\cal F}_1(\theta_c) -{\cal F}_2(\theta_c)]
\sin\theta_c\cos\phi_c\mbox{Re} \rho^0_{10}
\right\}  \ ,
\end{eqnarray}
\begin{eqnarray}
\label{w-1-cm}
W^1(\cos\theta_c,\phi_c, \rho^0)& = &
\frac{3}{8\pi}\frac{C^2p_0^2}{\sigma_0} \left\{ \sin^2\theta_c
\rho^1_{00} 
+ \frac{1}{2}\big[ {\cal F}_1^2(\theta_c)
+{\cal F}_2^2(\theta_c) \big] \rho^1_{11} 
\right.\nonumber\\
&&\left. + \sin^2\theta_c\cos 2\phi_c\rho^1_{1-1}
+\sqrt{2} [ {\cal F}_1(\theta_c) -{\cal F}_2(\theta_c)]
\sin\theta_c\cos\phi_c\mbox{Re} \rho^1_{10}
\right\}  \ ,
\end{eqnarray}
\begin{equation}
\label{w-2-cm}
W^2(\cos\theta_c,\phi_c, \rho^2) =
\frac{3}{8\pi}\frac{C^2p_0^2}{\sigma_0} \left\{ \sin^2\theta_c\sin 2\phi_c \mbox{Im}\rho^2_{1-1}
-\sqrt{2} [ {\cal F}_1(\theta_c) -{\cal F}_2(\theta_c)]
\sin \theta_c\sin\phi_c \mbox{Im}\rho^2_{10} \right\}
\end{equation}
and
\begin{equation}
\label{w-3-cm}
W^3(\cos\theta_c,\phi_c, \rho^2) =
\frac{3}{8\pi} \frac{C^2p_0^2}{\sigma_0}\left\{ \sin^2\theta_c\sin 2\phi_c \mbox{Im}\rho^3_{1-1}
-\sqrt{2} [ {\cal F}_1(\theta_c) -{\cal F}_2(\theta_c)]
\sin \theta_c\sin\phi_c \mbox{Im}\rho^3_{10} \right\} \ .
\end{equation}

Recalling again that $\theta_c$ is a function of $\theta$, 
strong kinematic correlations have been embedded in the above 
expressions. Naturally, one would expect that in the limit of $|{\bf q}|\to 0$,
 and hence $q_0\to M_V$, the azimuthal angles $(\theta_c, \phi_c)$
will be identical to $(\theta, \phi)$, and 
$C^2p_0^2/\sigma_0=p_0^2/|{\bf r}|^2=1$. 
These expressions then reduce to the ones derived 
in the vector-meson-rest frame~\cite{schilling}:
\begin{eqnarray}
\label{w-0-v}
W^0(\cos\theta,\phi, \rho^0)& = &
\frac{3}{8\pi}\left\{ \sin^2\theta
\rho^0_{00} + (1+\cos^2\theta)\rho^0_{11} + \sin^2\theta\cos 
2\phi\rho^0_{1-1}
\right.\nonumber\\
&&\left. +\sqrt{2}\sin 2\theta\cos\phi\mbox{Re} \rho^0_{10}
\right\}  \ ,
\end{eqnarray}
\begin{eqnarray}
\label{w-1-v}
W^1(\cos\theta,\phi, \rho^1)&=&
\frac{3}{8\pi}\left\{ \sin^2\theta
\rho^1_{00} + (1+\cos^2\theta)\rho^1_{11} + \sin^2\theta\cos 
2\phi\rho^1_{1-1}
\right.\nonumber\\
&&\left. +\sqrt{2}\sin 2\theta\cos\phi\mbox{Re} \rho^1_{10}
\right\}  \ ,
\end{eqnarray}
\begin{equation}
\label{w-2-v}
W^2(\cos\theta,\phi, \rho^2) =
\frac{3}{8\pi}\left\{ \sin^2\theta\sin 2\phi \mbox{Im}\rho^2_{1-1}
+\sqrt{2}\sin 2\theta\sin\phi \mbox{Im}\rho^2_{10} \right\}
\end{equation}
and
\begin{equation}
\label{w-3-v}
W^3(\cos\theta,\phi, \rho^3) =
\frac{3}{8\pi}\left\{ \sin^2\theta\sin 2\phi \mbox{Im}\rho^3_{1-1}
+\sqrt{2}\sin 2\theta\sin\phi \mbox{Im}\rho^3_{10} \right\} \ .
\end{equation}

As shown in Eq.~(\ref{m-trans}), the non-vanishing three momentum 
of the vector meson in the overall c.m. system 
introduces an additional term for the transverse 
vector meson decays. 
As a consequence, kinematic factors ${\cal F}_{1,2}(\theta_c)$, which 
are functions of the decay angle $\theta_c$, appear in the 
$TT$ and $TL$ decay distributions. Nevertheless, 
co-ordinate transformations also exist between $\theta_c$ and $\theta$. 
In comparison with 
Eqs.~(\ref{w-0-v})-(\ref{w-3-v}) for vector meson decay in its rest frame, 
kinematic factors in Eqs.~(\ref{w-0-cm})-(\ref{w-3-cm})
cannot be cleanly factorized out 
in the measurement of $V\to P\gamma$ in the overall c.m. system. 
This is different from the measurement of 
vector meson decay into spinless mesons~\cite{Kloet}, e.g. 
$\rho^0\to \pi^+\pi^-$ or $\phi\to K^+K^-$, where a kinematic factor 
can be taken out of the distribution functions by selecting the 
polar angle $\theta$ as the angle between the momentum difference 
of the final-state pions and the vector meson momentum in the overall 
c.m. system. 
Such a correlation also reflects the dynamical difference between 
$V\to P\gamma$ and $V\to PP$ from which we expect to learn 
more about the vector meson production mechanism via the measurement 
of the decay distributions.

In the next section, we will discuss some features arising from the kinematic 
correlations and their impact on the polarized beam asymmetry measurement 
in different frames.

\section{Kinematic correlations in polarized beam asymmetry}

In the production plane, 
the measurement of the linearly polarized beam asymmetry 
is defined as the cross section difference between 
polarizing the photons along
the $x_i$ ($\Phi=0^\circ$) and $y_i$-axis ($\Phi=90^\circ$),
which correspond to ${\bf P}_\gamma$ along $\mp x_i$, respectively.
The cross sections for these two polarizations thus can be expressed as:
\bea
\bar{W}_\perp(\Phi=90^\circ,\rho) &=&
\int_{\theta=0}^\pi\int_{\phi=0}^{2\pi} d\Omega W^0(\cos\theta_c, \phi_c, 
\rho^0)
-P_\gamma \int_{\theta=0}^\pi\int_{\phi=0}^{2\pi} d\Omega W^1(\cos\theta_c, 
\phi_c, \rho^1)
\nonumber\\
&=&\bar{W}^0(\rho^0)+P_\gamma \bar{W}^1(\rho^1) \ ,
\eea
and
\be
\bar{W}_\parallel(\Phi=0^\circ,\rho)
=\bar{W}^0(\rho^0)-P_\gamma \bar{W}^1(\rho^1) \ .
\ee
The linearly polarized photon asymmetry
is thus defined as:
\begin{equation}
\label{pola-photon-1}
\check{\Sigma}\equiv
\frac{\bar{W}_\perp(\Phi=90^\circ,\rho)-\bar{W}_\parallel(\Phi=0^\circ,\rho)}
{\bar{W}_\perp(\Phi=90^\circ,\rho)+\bar{W}_\parallel(\Phi=0^\circ,\rho)}
=P_\gamma \frac{\bar{W}^1(\rho^1)}{\bar{W}^0(\rho^0)} \ ,
\end{equation}
where $P_\gamma$ is determined by the experimental setup, 
and $\bar{W}^1(\rho^1)$ and $\bar{W}^0(\rho^0)$ are the integrals over 
$(\theta, \phi)$:
\bea
\bar{W}^1(\rho^1) & \equiv & 
\int d\Omega W^1(\cos\theta_c, \phi_c, \rho^1) \nonumber\\
&= &\frac{3}{8\pi}
\frac{1}{|{\bf r}|^2M_V^2}  \int d\Omega
\left\{ [ (p\cdot q)^2 -M_V^2 (p\cdot \epsilon_v^0)^2] \rho^1_{00}
+ [ (p\cdot q)^2 +M_V^2 (p\cdot \epsilon_v^0)^2] \rho^1_{11}\right\} ,
\eea
and
\bea
\bar{W}^0(\rho^0) & \equiv & 
\int d\Omega W^0(\cos\theta_c, \phi_c, \rho^0) \nonumber\\
&= &\frac{3}{8\pi}
\frac{1}{|{\bf r}|^2M_V^2}  \int d\Omega
\left\{ [ (p\cdot q)^2 -M_V^2 (p\cdot \epsilon_v^0)^2] \rho^0_{00}
+ [ (p\cdot q)^2 +M_V^2 (p\cdot \epsilon_v^0)^2] \rho^0_{11}\right\} ,
\eea
where the relation $p_0^2[{\cal F}_1^2(\theta_c)
+{\cal F}_2^2(\theta_c)]/2=[(p\cdot q)^2 + M_V^2(p\cdot \epsilon_v^0)^2]/M_V^2$ 
has been applied, and $\epsilon_v^0 =(|{\bf q}|, q_0 \hat{\bf q})/M_V $ 
is the longitudinal polarization vector of the vector meson. 

Note that the density matrix elements $\rho^\alpha$ ($\alpha=0$, 1, 2, 3) 
are independent of $\theta$ and $\phi$, we define the following two integrals: 
\bea
W_a & \equiv & \frac{3}{8\pi} \frac{1}{|{\bf r}|^2M_V^2}
\int d\Omega [(p\cdot q)^2- M_V^2 (p\cdot\epsilon_v^0)^2 ] \nonumber\\
W_b & \equiv & \frac{3}{8\pi} \frac{1}{|{\bf r}|^2M_V^2} 
\int d\Omega [(p\cdot q)^2+ M_V^2 (p\cdot\epsilon_v^0)^2 ] .
\eea
Hence, the polarized beam asymmetry can be expressed as
\be
\check{\Sigma} =
\frac{\rho^1_{00} + (W_b/W_a) \rho^1_{11}}
{\rho^0_{00} + (W_b/W_a) \rho^0_{11}} \ ,
\ee
and the ratio of the integrals gives:
\be
\frac{W_b}{W_a}= 2 + \frac 12 \frac{|{\bf q}|^2}{M_V^2}
\left(1+\frac 32\frac{M_P^2}{|{\bf r}|^2}\right) \ ,
\ee
where in the integration, the relation 
$2q_0 p_0=q_0^2 + p_0^2 -M_P^2 -({\bf r}-{\bf q}/2)^2$
has been used. 
In the vector-meson-rest frame, $|{\bf q}|\to 0$, we have 
$W_b/W_a=2$, and the polarized beam asymmetry reduces to 
\be
\label{ratio}
\check{\Sigma} =
\frac{\rho^1_{00} + 2 \rho^1_{11}}
{\rho^0_{00} + 2 \rho^0_{11}} \ ,
\ee
which is the familiar result derived in the vector meson decays 
into pseudoscalar mesons~\cite{zhao-omega}. 

Due to the non-trivial kinematic correlations arising from the 
decay distributions, the polarized photon beam asymmetry 
is also accompanied with a kinematic correlation factor 
when transforms from the vector-meson-rest frame to 
the overall c.m. frame. Although one, in principle, can 
select the vector-meson-rest frame as the working frame 
for deriving the polarization asymmetry, in reality, it will 
depend on how well the vector meson kinematic is reconstructed. 
For narrow states, such as $\omega$ and $\phi$, 
the transformation of the working 
frame from the overall c.m. one to the vector-meson-rest one 
should have relatively small ambiguities. However, for broad states, such as 
$\rho$, the sizeable width will lead to uncertainties 
of determing the three moment ${\bf q}$. The 
kinematic correlations hence will produce significant 
effects in the measurement of the polarization observables. 

This kind of situation makes 
Eq.~(\ref{ratio}) useful for evaluating the kinematic correlation effects. 
Firstly, note that the second term in Eq.~(\ref{ratio}) explicitly 
depends on the  mass and momentum of the pseudoscalar meson in 
the vector-meson-rest frame. It shows that the kinematic correlations 
will become more significant if the vector meson decays into a heavier 
pseudoscalar meson instead of a lighter one 
due to the increasing ratio $M_P^2/|{\bf r}|^2$. 
For instance, for the final state decays of  
$\omega\to\eta\gamma$, $M_\pi^2/|{\bf r}_\pi|^2=0.13$, 
 and for $\omega\to \pi^0\gamma$, 
$M_\eta^2/|{\bf r}_\eta|^2=7.56$. This ratio is essentially a constant for 
a fixed decay channel. 
Therefore, Eq.~(\ref{ratio}) suggests that 
above the vector meson production channel and at a fixed production energy, 
kinematic correlation effects 
should show up increasingly significant for heavier pseudoscalar decay channels.

Secondly, for a fixed pseudoscalar decay channel, 
Eq.~(\ref{ratio}) shows that the kinematic correlation should also 
become increasingly important with the increasing reaction energies
due to the term proportional to $|{\bf q}|^2/M_V^2$. 
In another word, only when the vector meson is produced near threshold, 
i.e. $|{\bf q}|^2/M_V^2 << 1$, can the kinematic correlation effects 
be neglected as a reasonable approximation. Otherwise, 
uncertainties from such correlations should be cautioned.

\section{Summary}

We have carried out an analysis of vector meson decay into a pseudoscalar meson 
plus a photon in photoproduction reactions, e.g.
$\omega\to \pi^0\gamma$, $\phi\to \eta\gamma$, $\rho\to \pi\gamma$, and $K^*\to K\gamma$, etc.
Comparing with the previous measurements of vector meson decays into 
pseudoscalars, this channel has advantages of providing additional 
dynamical information about the vector meson production mechanism 
due to the different spin structures carried by the 
differential decay distributions. 

It was found that kinematic correlations would become important 
in this decay channel in the overall c.m. system. 
An explicit relation for the correlations 
between the vector-meson-rest frame and the overall c.m. system was derived in Eq.(33), 
which highlighted the kinematic sensitivities of the polarization observables  
studied in different reactions and different pseudoscalar meson decay channels. 
It is thus useful for providing guidance for experimental investigation
of vector meson photoproduction for the purpose of studying nucleon
resonance excitations in polarization reactions.

Although
the decay channel $V\to P\gamma$ generally has small branching ratios
for most vector mesons, it is relatively large
for the omega meson with $b_{\omega\to\pi^0\gamma}=(8.5\pm 0.5)\% $~\cite{pdg2004}.
Therefore, for $\omega$ meson photoproduction, this channel 
will not only increase the experimental statistics, but also 
provide an independent measurement of polarized beam asymmetry 
which can be then compared with the one measured 
in $\omega\to \pi^0\pi^+\pi^-$. 
For $\phi$ and $\rho^0$, their dominant decays are into two pseudoscalars. 
Hence, additional spin structure information can be expected in  
$\phi\to\eta\gamma$ and $\rho^0\to \pi^0\gamma$. 
In particular, $br_{\phi\to\eta\gamma}=(1.295\pm 0.025)\% $~\cite{pdg2004} 
is still sizeable, and makes this channel an important source for the $\phi$ 
meson photoproduction mechanism near threshold. 
Experimental facilities at ESRF (GRAAL), 
SPring-8 (LEPS) and ELSA (Crystal Ball) with linearly polarized photon 
beams and charge-neutral particle detectors should have advantages 
for addressing this issue.

We also derived the single polarization observables 
for vector meson photoproduction
in terms of the density matrix elements
for the polarized particles. 
Those elements can be directly related to the experimental 
measurement of the corresponding angular distributions
of the vector meson decays into either spinless mesons
or a spinless meson plus a photon. For linearly polarized photon beams
the angular distribution of the photoproduced-vector-meson decay 
into spinless particles has been developed in the literature~\cite{schilling}. 
The  polarization observables in terms of the bilinear helicity product 
of the helicity amplitudes have also been discussed widely 
in the literature~\cite{tabakin}. In the Appendix, we presented   
the vector meson decay distribution functions in terms of those 
measurable density matrix elements defined for those polarized particles. 
This should be useful for future experimental analyses 
at GRAAL, JLab, and SPring-8, 
with polarized targets, recoil polarization or beam-target 
double polarizations.

\section*{Acknowledgement}

This work is supported by the U.K. Engineering and Physical
Sciences Research Council Advanced Fellowship (Grant No. GR/S99433/01), 
and the NSF grant PHY-0417679. 
Q.Z. thanks A. D'Angelo, J.-P. Didelez, E. Hourany, D. Rebreyend 
for many useful discussions about the GRAAL experiment.

\section*{Appendix: Density matrix elements for polarization observables}

In Ref.~\cite{schilling}, the density matrix elements for
polarized photon beams were derived. We shall adopt the same
convention and
derive density matrix elements for other single polarization
(i.e. polarized target, recoil polarization, and
vector meson polarization), and double polarization observables.
First of all, we will outline several basic aspects
of density matrix elements with polarized photon beams, which 
will be useful for further discussions.

For the convenience of comparing with other analyses~\cite{tabakin},
we also express the invariant amplitudes as
the following 12 independent helicity amplitudes:
\begin{eqnarray}
\label{helicity}
H_{1\lambda_v}&=  &\langle \lambda_v, \lambda_f=+1/2|T|\lambda_\gamma=1,
\lambda_i=-1/2\rangle\nonumber\\
H_{2\lambda_v}&=  &\langle \lambda_v, \lambda_f=+1/2|T|\lambda_\gamma=1,
\lambda_i=+1/2\rangle\nonumber\\
H_{3\lambda_v}&=  &\langle \lambda_v, \lambda_f=-1/2|T|\lambda_\gamma=1,
\lambda_i=-1/2\rangle\nonumber\\
H_{4\lambda_v}&=  &\langle \lambda_v, \lambda_f=-1/2|T|\lambda_\gamma=1,
\lambda_i=+1/2\rangle \ .
\end{eqnarray}

\subsection{Convention and kinematics}

The general form of the angular distribution for the vector meson decay
into spinless particles (e.g., $\omega\to\pi^+\pi^-\pi^0$,
$\rho^0\to\pi^+\pi^-$, $\phi\to K^+ K^-$) is
\begin{equation}
\label{distr}
\frac{d N}{d\cos\theta d\phi}=W(\cos\theta,\phi)=
\sum_{\lambda_v\lambda_v^\prime}
\langle\theta,\phi | M |\lambda_v\rangle \rho_{\lambda_v\lambda_v^\prime}(V)
\langle \lambda_v^\prime| M^\dag |\theta,\phi\rangle \ ,
\end{equation}
where $M$ is the vector meson decay amplitude and can be factorized as
\begin{equation}
\langle\theta,\phi | M |\lambda_v\rangle=C\sqrt{\frac{3}{4\pi}}
D^{1*}_{\lambda_v 0}(\phi,\theta,-\phi) \ ;
\end{equation}
The decay angles, $\theta$ and $\phi$, are defined as
the polar and azimuthal angles of the flight direction of one of
the decay particles in the vector meson rest frame in the case of
two-body decay, e.g. $\rho^0\to \pi^+\pi^-$, $\phi\to K^+ K^-$.
For three-body decay of the vector meson, e.g. $\omega\to \pi^0\pi^+\pi^-$,
$\theta$ and $\phi$ denote the polar and  azimuthal angles
of the normal direction of the decay plane.
The constant $C$ is independent of $\lambda_v$
and determined by the vector meson decay width;
the Wigner rotation functions $D$ follow the convention
of Ref.~\cite{rose} as defined by Eq.~(\ref{rotation}). 

Consequently, the angular distribution of Eq.~(\ref{distr})
can be expressed in terms of the vector meson density matrices
$\rho_{\lambda_v\lambda_v^\prime}(V)$:
\begin{equation}
\label{distr-general}
W(\cos\theta,\phi)=\frac{3}{4\pi}\sum_{\lambda_v\lambda_v^\prime}
D^{1*}_{\lambda_v 0}(\phi,\theta,-\phi)
\rho_{\lambda_v\lambda_v^\prime}(V)
D^1_{\lambda_v^\prime 0}(\phi,\theta,-\phi) \ ,
\end{equation}
where $\rho(V)$ is Hermitian,
$\rho_{\lambda_v\lambda_v^\prime}(V)=\rho^*_{\lambda_v^\prime\lambda_v}(V)$.
Meanwhile, since $W(\cos\theta,\phi)$ is a linear function of $\rho(V)$,
it can be decomposed into a linear combination in terms of the polarization
status of the particles.

The above equation is general for the vector meson decays
into spinless particles.
The vector meson density matrix element $\rho(V)$
can be related to the polarization status of the initial and final state
particles in their spin space via the transition amplitudes
$T_{\lambda_v\lambda_f,\lambda_\gamma\lambda_i}$.
Therefore, for different polarization reactions with the vector meson
decays into spinless particles, $\rho(V)$ is the source containing
information about the polarization observables.
The angular distribution, as a function of $\rho(V)$, hence provides
access to the physics reflected by $\rho(V)$.
In experiment, $\rho(V)$ is a quantity that can be derived 
from the data for vector meson decay distributions,
while in theory, it can be calculated with dynamical 
models~\cite{zhao-98,zhao-omega,Friman:1995qm,Zhao:2001jw,Zhao:2001ue,Titov:2002zy,Titov:2002iv,Oh:2000zi,Titov:1998bw,Oh:2002rb,Oh:2001bq,Shklyar:2004ba,Penner:2002ma,Feuster:1998cj}.

Since $\rho(V)$ is related to the polarization of the initial
and final state particles in their spin space,
this is equivalent to saying that for different polarizations of the initial
and final state particles there are different density matrix elements
can be measured. In this sense, we are also interested in the number
of minimum measurements which can provide the maximum amount of
information on the transition mechanisms~\cite{tabakin}.

As follows, we will first re-derive the polarized beam asymmetry
in terms of the density matrix elements following Ref.~\cite{schilling},
and then derive other polarization asymmetries
in terms of the corresponding density matrix elements.

\subsection{Polarized beam asymmetry}

Substituting Eq.~(\ref{photon-pol}) into (\ref{distr-general}),
we can easily reproduce the results of Ref.~\cite{schilling} and obtain
the angular distribution:
\begin{eqnarray}
\label{decay-distribution}
W(\cos \theta, \phi, \Phi) &=
& W^0(\cos\theta, \phi, \rho^0_{\lambda_v\lambda_v^\prime})
- P_\gamma \cos 2\Phi
W^1(\cos\theta,\phi,\rho^1_{\lambda_v\lambda_v^\prime}) \nonumber\\
& & -P_\gamma \sin 2\Phi
W^2(\cos\theta,\phi,\rho^2_{\lambda_v\lambda_v^\prime})\nonumber\\
& & + \lambda_\gamma P_\gamma
W^3(\cos\theta,\phi,\rho^3_{\lambda_v\lambda_v^\prime}) \ .
\end{eqnarray}
where
\begin{eqnarray}
\label{dd-un}
W^0(\cos\theta, \phi, \rho^0)&=&
\frac{3}{4\pi} [\frac 12 \sin^2\theta + \frac 12 (3\cos^2\theta -1)
\rho^0_{00}\nonumber\\
&&-\sqrt{2} \mbox{Re} \rho^0_{10}\sin 2\theta\cos\phi
-\rho^0_{1-1}\sin^2\theta\cos 2\phi] \ , \nonumber \\
W^1(\cos\theta,\phi,\rho^1)&=&
\frac{3}{4\pi} [\rho^1_{11} \sin^2\theta + 
\rho^1_{00}\cos^2\theta\nonumber\\
&&-\sqrt{2} \mbox{Re} \rho^1_{10}\sin 2\theta\cos\phi
-\rho^1_{1-1}\sin^2\theta\cos 2\phi] \ , \nonumber\\
W^2(\cos\theta,\phi,\rho^2)&=&
\frac{3}{4\pi} [\sqrt{2} \mbox{Im} \rho^2_{10}\sin 2\theta\sin\phi
+ \mbox{Im}\rho^2_{1-1}\sin^2\theta\sin 2\phi ] \ , \nonumber\\
W^3(\cos\theta,\phi,\rho^3) &=&
\frac{3}{4\pi} [ \sqrt{2} \mbox{Re}\rho^3_{10} \sin 2\theta\sin\phi
+ \mbox{Im} \rho^3_{1-1} \sin ^2\theta\sin 2\phi ] \ .
\end{eqnarray}

For the polarized photon beams, the density matrix element
$\rho^0$ corresponds to the unpolarized photon measurement,
$\rho^1$ and $\rho^2$ correspond to the linearly polarized photon,
while $\rho^3$ to the circularly polarized photon.
The linearly polarized beam asymmetry then is given 
by the cross section differences between
polarizing the photons along
the $x_i$ ($\Phi=0^\circ$) and $y_i$-axis ($\Phi=90^\circ$),
which correspond to ${\bf P}_\gamma$ along $\mp x_i$, respectively.
Similar to Eq.~(\ref{pola-photon-1}), we have
\begin{equation}
\label{pola-photon}
\check{\Sigma}\equiv
\frac{\bar{W}_\perp(\Phi=90^\circ,\rho)-\bar{W}_\parallel(\Phi=0^\circ,\rho)}
{\bar{W}_\perp(\Phi=90^\circ,\rho)+\bar{W}_\parallel(\Phi=0^\circ,\rho)}
=P_\gamma \frac{\rho^1_{00}+2\rho^1_{11}}{\rho^0_{00}+2\rho^0_{11}} \ .
\end{equation}

In terms of the helicity amplitudes the above expression can be
written as
\begin{eqnarray}
\check{\Sigma}
&=&\frac 12\{ -H^r_{1-1}H^r_{41}-H^i_{1-1}H^i_{41}
+H^r_{10}H^r_{40}+H^i_{10}H^i_{40}\nonumber\\
&&-H^r_{11}H^r_{4-1}-H^i_{11}H^i_{4-1}
+H^r_{2-1}H^r_{31}+H^i_{2-1}H^i_{31}\nonumber\\
&&-H^r_{20}H^r_{30}-H^i_{20}H^i_{30}
+H^r_{21}H^r_{3-1}+H^i_{21}H^i_{3-1} \} \nonumber\\
&=&\frac 12\langle H|\Gamma^4\omega^A|H \rangle \ .
\end{eqnarray}
The above expression is the same as that defined in Ref.~\cite{tabakin}.

\subsection{Target polarization asymmetry}

Proceeding to other single polarization observables,
we shall derive the angular distributions
of the vector meson decays into spinless particles
and express the  polarization observables in terms of the
corresponding density matrix elements.

The polarization status of the target is defined as
\begin{equation}
\rho(N)=\frac{1}{2}(I_{N_i}+\vsig\cdot {\bf P}_{N_i}) \ ,
\end{equation}
where $I_{N_i}$ is the $2\times 2$ unity matrix in the spin space of the 
target;
${\bf P}_{N_i}$ is the polarization vector for the target nucleon.
In the coordinates of the initial states, we define
${\bf P}_{N_i}\equiv P_{N_i}(\cos\Phi, \ \sin\Phi, \ 0)$
for polarizing the initial nucleon within the $x_i$-$y_i$ plane,
and  ${\bf P}_{N_i}\equiv P_{N_i}(0, \ 0, \ \lambda_i)$
for polarizing the initial
nucleon along $z_i$,
where $P_{N_i}$ is the initial-nucleon degree of polarization, and
$\Phi$ is the angle between the polarization vector of the initial nucleon
and the production plane ($x_i$-$z_i$ plane).

The polarized target density matrix elements can be related to the elements
of the vector meson decay via the production amplitudes $T$:
\begin{equation}
\rho_{\lambda_v\lambda_v^\prime}(V)=
\frac{1}{N} \sum_{\lambda_f\lambda_\gamma\lambda_i\lambda_i^\prime}
T_{\lambda_v\lambda_f,\lambda_\gamma\lambda_i}
\rho_{\lambda_i\lambda_i^\prime}(N_i)
T_{\lambda_v^\prime\lambda_f,\lambda_\gamma\lambda_i^\prime}^* \ .
\end{equation}

The decomposition of the polarization status of the target gives
\begin{equation}
\rho(V)=\rho^0 +\sum_{\alpha=1}^3 P_{N_i}^\alpha \rho^\alpha \ ,
\end{equation}
where $\rho^0$ denotes the density matrix elements with the unpolarized
target, and $\rho^{1,2,3}$ denotes the elements with the target polarized
along $x_i$, $y_i$ and $z_i$-axis in the initial frame.
Explicitly, the elements are given as
\begin{eqnarray}
\label{dens-target}
\rho^0_{\lambda_v\lambda_v^\prime}&=&\frac{1}{2N}
\sum_{\lambda_f\lambda_\gamma\lambda_i}
T_{\lambda_v\lambda_f,\lambda_\gamma\lambda_i}
T^*_{\lambda_v^\prime\lambda_f,\lambda_\gamma\lambda_i} \nonumber\\
\rho^1_{\lambda_v\lambda_v^\prime}&=&\frac{1}{2N}
\sum_{\lambda_f\lambda_\gamma\lambda_i}
T_{\lambda_v\lambda_f,\lambda_\gamma-\lambda_i}
T^*_{\lambda_v^\prime\lambda_f,\lambda_\gamma\lambda_i} \nonumber\\
\rho^2_{\lambda_v\lambda_v^\prime}&=&\frac{i}{2N}
\sum_{\lambda_f\lambda_\gamma\lambda_i}\hat{\lambda}_i
T_{\lambda_v\lambda_f,\lambda_\gamma-\lambda_i}
T^*_{\lambda_v^\prime\lambda_f,\lambda_\gamma\lambda_i} \nonumber\\
\rho^3_{\lambda_v\lambda_v^\prime}&=&\frac{1}{2N}
\sum_{\lambda_f\lambda_\gamma\lambda_i}\hat{\lambda}_i
T_{\lambda_v\lambda_f,\lambda_\gamma\lambda_i}
T^*_{\lambda_v^\prime\lambda_f,\lambda_\gamma\lambda_i} \ ,
\end{eqnarray}
where $\hat{\lambda}_i=\pm 1$ represents the sign of $\lambda_i$.

The vector meson decay distribution depends on the properties
of the density matrix elements defined above.
Applying the parity conservation relation and the requirement
of the density matrix elements to be Hermitian,
we obtain the matrix elements
for $\rho^\alpha_{\lambda_v\lambda_v^\prime}$ with $\alpha=1$, 2, 3,
corresponding to the polarization of the initial nucleon spin along
$x_i$, $y_i$, and $z_i$.
Substituting the elements into Eq.~(\ref{distr-general}),
we obtain the vector meson distribution in terms of the
different polarization status of the target nucleons:
\begin{eqnarray}
\label{pol-t-distr}
W(\cos \theta, \phi, \Phi) &=
& W^0(\cos\theta, \phi, \rho^0_{\lambda_v\lambda_v^\prime})
+ P_{N_i} \cos \Phi
W^1(\cos\theta,\phi,\rho^1_{\lambda_v\lambda_v^\prime}) \nonumber\\
& & +P_{N_i} \sin \Phi
W^2(\cos\theta,\phi,\rho^2_{\lambda_v\lambda_v^\prime})\nonumber\\
& & + \hat{\lambda_i} P_{N_i}
W^3(\cos\theta,\phi,\rho^3_{\lambda_v\lambda_v^\prime}) \ .
\end{eqnarray}
where
\begin{eqnarray}
\label{w-func}
W^0(\cos\theta, \phi, \rho^0)&=&
\frac{3}{4\pi} [\frac 12 \sin^2\theta + \frac 12 (3\cos^2\theta -1)
\rho^0_{00}\nonumber\\
&&-\sqrt{2} \mbox{Re} \rho^0_{10}\sin 2\theta\cos\phi
-\rho^0_{1-1}\sin^2\theta\cos 2\phi] \ , \nonumber \\
W^1(\cos\theta,\phi,\rho^1)&=&
\frac{3}{4\pi} [\sqrt{2} \mbox{Im} \rho^1_{10}\sin 2\theta\sin\phi
+ \mbox{Im}\rho^1_{1-1}\sin^2\theta\sin 2\phi ] \ , \nonumber\\
W^2(\cos\theta,\phi,\rho^2)&=&
\frac{3}{4\pi} [\rho^2_{11} \sin^2\theta + 
\rho^2_{00}\cos^2\theta\nonumber\\
&&-\sqrt{2} \mbox{Re} \rho^2_{10}\sin 2\theta\cos\phi
-\rho^2_{1-1}\sin^2\theta\cos 2\phi] \ , \nonumber\\
W^3(\cos\theta,\phi,\rho^3) &=&
\frac{3}{4\pi} [ \sqrt{2} \mbox{Re}\rho^3_{10} \sin 2\theta\sin\phi
+ \mbox{Im} \rho^3_{1-1} \sin ^2\theta\sin 2\phi ] \ ,
\end{eqnarray}
where $\Phi$ again denotes the angle between the polarization vector
of the initial nucleon and the production plane.

As addressed earlier,
the corresponding angular distribution of the vector meson decays
into pions with the polarized target
has the same form as Eq.~(\ref{distr-general}), while the density
matrix elements represent different dynamical information.
In analogy with the polarized photon measurement,
the cross sections for polarizing the initial nucleon spin projection
along $\pm y_i$ axis, i.e. $\Phi=\pm 90^\circ$, can be obtained
by summing all the events together:
\bea
\bar{W}_\uparrow(\Phi=90^\circ,\rho) &=&
\int_{\theta=0}^\pi\int_{\phi=0}^{2\pi} d\Omega W^0(\cos\theta, \phi, 
\rho^0)
+P_{N_i} \int_{\theta=0}^\pi\int_{\phi=0}^{2\pi} d\Omega W^2(\cos\theta, 
\phi, \rho^2)
\nonumber\\
&=&\bar{W}^0(\rho^0)+P_{N_i} \bar{W}^2(\rho^2) \ ,
\eea
and
\be
\bar{W}_\downarrow(\Phi=-90^\circ,\rho)
=\bar{W}^0(\rho^0)-P_{N_i} \bar{W}^2(\rho^2) \ .
\ee
The polarized target asymmetry is hence defined as the cross section
differences between these two polarizations:
\begin{equation}
\label{pola-target}
\check{T}\equiv
\frac{\bar{W}_\uparrow(\Phi=90^\circ,\rho)-\bar{W}_\downarrow(\Phi=-90^\circ,\rho)}
{\bar{W}_\uparrow(\Phi=90^\circ,\rho)+\bar{W}_\downarrow(\Phi=-90^\circ,\rho)}
=P_{N_i}\frac{\rho^2_{00}+2\rho^2_{11}}{\rho^0_{00}+2\rho^0_{11}} \ ,
\end{equation}
where $P_{N_i}$ only depends on the experimental setup.

Comparing with the polarized beam asymmetry,
it shows that the density matrix elements $\rho^{1,2,3}$
defined in the target polarization
have different properties. In particular, the asymmetry
of polarizing the initial nucleon along $x_i$ axis would be zero
if the vector meson decay events are integrated over $\theta$ and $\phi$.

The dynamic significance can be seen more transparently
in terms of the helicity amplitudes, i.e.
\begin{eqnarray}
\check{T}
&=&\sum_{\lambda_v} \{ H^r_{1\lambda_v} H^i_{2\lambda_v}
-H^i_{1\lambda_v} H^r_{2\lambda_v}
+H^r_{3\lambda_v} H^i_{4\lambda_v}
-H^i_{3\lambda_v} H^r_{4\lambda_v} \}\nonumber\\
&=&-\frac 12\langle H|\Gamma^{10} \omega^1|H \rangle \ ,
\end{eqnarray}
which now involves interferences between different helicity amplitudes,
and again arrives at the same result as that of Ref.~\cite{tabakin}.

\subsection{Recoil polarization asymmetry}

Similarly, we can also  establish the relations
for the density matrix elements between the vector meson and the recoil
nucleon by defining
\begin{equation}
\rho(N_f)=\frac{1}{2}(I_{N_f}+\vsig\cdot {\bf P}_{N_f}) \ ,
\end{equation}
where $I_{N_f}$ is the $2\times 2$ unity matrix
in the spin space of the recoil nucleon and
${\bf P}_{N_f}$ is the polarization direction
defined in the final state coordinates, i.e. the frame of
$(x_f, y_f, z_f)$ in Fig.~\ref{fig:(1)}.
In the spin space for the recoil nucleon,
the decomposition of the recoil polarizations leads to:
\begin{eqnarray}
\label{dens-recoil}
\rho^0_{\lambda_v\lambda_v^\prime}&=&\frac{1}{2N}
\sum_{\lambda_f\lambda_\gamma\lambda_i}
T_{\lambda_v\lambda_f,\lambda_\gamma\lambda_i}
T^*_{\lambda_v^\prime\lambda_f,\lambda_\gamma\lambda_i} \nonumber\\
\rho^1_{\lambda_v\lambda_v^\prime}&=&\frac{1}{2N}
\sum_{\lambda_f\lambda_\gamma\lambda_i}
T_{\lambda_v-\lambda_f,\lambda_\gamma\lambda_i}
T^*_{\lambda_v^\prime\lambda_f,\lambda_\gamma\lambda_i} \nonumber\\
\rho^2_{\lambda_v\lambda_v^\prime}&=&\frac{i}{2N}
\sum_{\lambda_f\lambda_\gamma\lambda_i}\hat{\lambda}_f
T_{\lambda_v-\lambda_f,\lambda_\gamma\lambda_i}
T^*_{\lambda_v^\prime\lambda_f,\lambda_\gamma\lambda_i} \nonumber\\
\rho^3_{\lambda_v\lambda_v^\prime}&=&\frac{1}{2N}
\sum_{\lambda_f\lambda_\gamma\lambda_i}\hat{\lambda}_f
T_{\lambda_v\lambda_f,\lambda_\gamma\lambda_i}
T^*_{\lambda_v^\prime\lambda_f,\lambda_\gamma\lambda_i} \ ,
\end{eqnarray}
where $\hat{\lambda}_f=\pm 1$ represents the sign of $\lambda_f$.

The angular distribution of the vector meson decays can then be expressed as
\begin{eqnarray}
\label{pol-r-distr}
W(\cos \theta, \phi, \Phi) &=
& W^0(\cos\theta, \phi, \rho^0_{\lambda_v\lambda_v^\prime})
+ P_{N_f} \cos \Phi
W^1(\cos\theta,\phi,\rho^1_{\lambda_v\lambda_v^\prime}) \nonumber\\
& & +P_{N_f} \sin \Phi
W^2(\cos\theta,\phi,\rho^2_{\lambda_v\lambda_v^\prime})\nonumber\\
& & + \hat{\lambda_f} P_{N_f}
W^3(\cos\theta,\phi,\rho^3_{\lambda_v\lambda_v^\prime}) \ .
\end{eqnarray}
where the $W^{0,1,2,3}$ corresponds to distributions with
the final state baryon unpolarized, and polarized along
$x_f$, $y_f$, and $z_f$ directions. The expressions of $W^{0,1,2,3}$
are the same as Eq.~(\ref{w-func}).

Similar to the derivation in former subsections,
the recoil polarization asymmetry
is defined as the cross section differences between polarizing the
final-state nucleon along $\pm y_f$ axis (corresponding to
$\Phi=\pm 90^\circ$):
\begin{equation}
\check{P}_{N_f}\equiv 
\frac{\bar{W}_\uparrow(\Phi=90^\circ,\rho)-\bar{W}_\downarrow(\Phi=-90^\circ,\rho)}
{\bar{W}_\uparrow(\Phi=90^\circ,\rho)+\bar{W}_\downarrow(\Phi=-90^\circ,\rho)}
=P_{N_f}\frac{\rho^2_{00}+2\rho^2_{11}}
{\rho^0_{00}+2\rho^0_{11}} \ ,
\end{equation}
which is the familiar form as Eq.~(\ref{pola-target}).

Again, in terms of the helicity amplitudes,
the dynamic significance can be seen:
\begin{eqnarray}
\check{P}_{N_f}
&=&\sum_{\lambda_v} \{H^i_{3\lambda_v}H^r_{1\lambda_v}
-H^r_{3\lambda_v}H^i_{1\lambda_v}
+H^i_{4\lambda_v}H^r_{2\lambda_v}
-H^r_{4\lambda_v}H^i_{2\lambda_v} \} \nonumber\\
&=&\frac 12\langle H|\Gamma^{12}\omega^1|H\rangle \ ,
\end{eqnarray}
which is the same as that given by Ref.~\cite{tabakin}.

\subsection{Beam-target double polarization asymmetry}

With the density matrix elements for the photon beams, target nucleon,
recoil nucleon, and vector meson, one can proceed to the investigation
of double polarizations in terms of the vector meson density matrix 
elements.

We shall start with the beam-target double polarization, of which
experiments are to be available at GRAAL~\cite{didelez} and 
JLab~\cite{cole}.
Similar to the single polarizations, the vector meson density matrix 
elements
is related to the polarization status of the photon and target nucleon.
Namely, we have
\begin{eqnarray}
\rho(V)&=& T\rho(\gamma)\rho(N_i) T^\dag\nonumber\\
&=&\frac 14 T(I_\gamma +\vsig\cdot{\bf P}_\gamma)
(I_{N_i}+\vsig\cdot{\bf P}_{N_i}) T^\dag \nonumber\\
&=&\frac 14 T (I_\gamma I_{N_i} +\vsig\cdot{\bf P}_\gamma I_{N_i}
+I_\gamma \vsig\cdot{\bf P}_{N_i}
+\vsig\cdot{\bf P}_\gamma\vsig\cdot{\bf P}_{N_i})
T^\dag \ ,
\end{eqnarray}
where one can see that the double polarization only involves the
last term in the bracket. The above formula reads that the double 
polarization
eventually provides access to the single polarizations as well.

We introduce the index $\alpha$ and $\beta$
for the polarization status of the photon beams
and the target nucleon, and express the B-T polarization as
\begin{eqnarray}
W(\cos\theta,\phi,\rho^{\alpha\beta}) &=&
W^{00}(\cos\theta,\phi,\rho^{00}) \nonumber\\
&+&\sum_{\alpha=1}^{3} P_\gamma^\alpha
W^{\alpha 0}(\cos\theta,\phi,\rho^{\alpha 0})\nonumber\\
&+& \sum_{\beta=1}^{3} P_{N_i}^\beta
W^{0\beta}(\cos\theta,\phi,\rho^{0\beta})\nonumber\\
&+& \sum_{\alpha,\beta=1}^{3} P_\gamma^\alpha P_{N_i}^\beta
W^{\alpha\beta}(\cos\theta,\phi,\rho^{\alpha\beta}) \ .
\end{eqnarray}
We shall concentrate on the last term, where both the photon and
the target nucleon are polarized.
The density matrix elements thus can be expressed as
\begin{equation}
\rho^{\alpha\beta}_{\lambda_v\lambda_v^\prime}(V)=
\frac{1}{N} 
\sum_{\lambda_f\lambda_\gamma\lambda_\gamma^\prime\lambda_i\lambda_i^\prime}
T_{\lambda_v\lambda_f,\lambda_\gamma\lambda_i}
\rho^\alpha_{\lambda_\gamma\lambda_\gamma^\prime}(\gamma)
\rho^\beta_{\lambda_i\lambda_i^\prime}(N_i)
T_{\lambda_v^\prime\lambda_f,\lambda_\gamma^\prime\lambda_i^\prime}^* \ .
\end{equation}
Given the polarization status of the photon and target nucleon,
one can derived the transition elements
$\rho^\alpha_{\lambda_\gamma\lambda_\gamma^\prime}(\gamma)$
and $\rho^\beta_{\lambda_i\lambda_i^\prime}(N_i)$ in their spin space.
For instance,
for the polarization that the photon is linearly polarized along
$x$-axis ($\alpha=1$) and the target nucleon is polarized along
$y$-axis ($\beta=2$), we have
\begin{equation}
\rho^{12}_{\lambda_v\lambda_v^\prime}(V)=
\frac{i}{4N}\sum_{\lambda_f\lambda_\gamma\lambda_i}
\hat{\lambda}_i
T_{\lambda_v\lambda_f,-\lambda_\gamma-\lambda_i}
T_{\lambda_v^\prime\lambda_f,\lambda_\gamma\lambda_i} \ ,
\end{equation}
where $\hat{\lambda}_i$ denotes the sign of $\lambda_i$.

The angular distribution again gives access to the experimental measurement
of the B-T asymmetry in terms of the density matrix elements:
\begin{equation}
{\cal C}^{\gamma N_i}_{xy}=\frac{\rho^{12}_{00}+2\rho^{12}_{11}}
{\rho^{00}_{00}+2\rho^{00}_{11}} \ ,
\end{equation}
which has the same form as other single polarizations;
$\rho^{00}_{00}$ and $\rho^{00}_{11}$
denote the unpolarized density matrix elements
and $\rho^{00}_{00}+2\rho^{00}_{11}=1$ due to the normalization.
Undoubtedly, the dynamic information contains in the helicity products
selected by the polarization status, and again,
the B-T polarization experiment will pick up
the double polarization asymmetry:
\begin{eqnarray}
{\cal C}^{\gamma N_i}_{xy}&=&-\frac{i}{2}
\{  H^*_{11}H_{3-1} + H^*_{21} H_{4-1} -H^*_{31} H_{1-1} -H^*_{41} H_{2-1}
\nonumber\\
&& +H^*_{1-1} H_{31} + H^*_{2-1} H_{41} - H^*_{3-1} H_{11} - H^*_{4-1} 
H_{21}
\nonumber\\
&& - H^*_{10} H_{30} - H^*_{20} H_{40} + H^*_{30} H_{10} + H^*_{40} H_{20} 
\}
\nonumber\\
&=& -\frac 12 \sum_{a,b,\lambda_v,\lambda_v^\prime}
H^*_{a\lambda_v}\Gamma^{12}_{ab}\omega^A_{\lambda_v\lambda_v^\prime}
H_{b\lambda_v^\prime} \ ,
\end{eqnarray}
which is the same as defined in Ref.~\cite{tabakin} and
$\Gamma^{12}_{ab}$ and $\omega^A_{\lambda_v\lambda_v^\prime}$
are $4\times 4$ and $3\times 3$ Hermitian matrices.

\end{document}